\documentclass[usenatbib]{aa}  

\usepackage{graphicx}
\usepackage{multirow}
\usepackage{rotating}
\usepackage{colortbl}
\usepackage{color}
\usepackage[fleqn]{amsmath}
\usepackage{amssymb}
\usepackage{amsfonts}
\usepackage{verbatim}
\usepackage{scalefnt}
\usepackage[percent]{overpic}
\usepackage{ulem}
\usepackage[hidelinks,colorlinks=true,linkcolor=blue,citecolor=blue]{hyperref}
\usepackage{cleveref}
\usepackage{url}

\usepackage{float}
\usepackage{txfonts}

\begin{document}

   \title{Gravitational Wave Memory Imprints on the CMB from Populations \\ of Massive Black Hole Mergers}

   \subtitle{}

   \author{Lorenz Zwick
          \inst{1} \and David O'Neill\inst{1} \and Kai Hendriks\inst{1} \and Philip Kirkeberg\inst{1} \and Miquel Miravet-Tenés\inst{2}}

   \institute{Niels Bohr International Academy, Niels Bohr Institute, Blegdamsvej 17, DK-2100 Copenhagen Ø, Denmark\\
              \email{lorenz.zwick96@gmail.com}
         \and
             Departament d'Astronomia i Astrofísica, Universitat de València, Dr Moliner 50, 46100, Burjassot (València), Spain}

   \date{Received April 10, 2024; accepted ???}

 
  \abstract
   {}
   {To showcase and characterise the rich phenomenology of temperature fluctuation patterns that are imprinted on the CMB by the gravitational wave memory (GWM) of massive black hole (BH) mergers.}
   {We analyse both individual binaries as well as populations of binaries, distributed in local cosmological boxes at a given redshift.}
   {The magnitude of the temperature fluctuations scales primarily as a function of binary total mass and pattern angular scale, and accumulates as a random-walk process when populations of mergers are considered. Fluctuations of {order $\sim 10^{-12}$ K are} reached across scales of $\sim 1'$ to $\sim 1^{\circ}$ for realistic volumetric merger rates of 10$^{-3}$ Mpc$^{-3}$ Gyr$^{-1}$, as appropriate for massive galaxies at $z=1$. We determine numerically that GWM temperature fluctuations result in a universal power spectrum with a scaling of $P(k)\propto k^{-2.7}$.}
   {While not detectable given the limitations of current all-sky CMB surveys, our work explicitly shows how every black hole merger in the Universe left us its unique faint signature.}

   \keywords{Black hole physics -- gravitational waves -- cosmic background radiation}
   \maketitle
%
\section{Introduction}
Gravitational wave memory (GWM) is a phenomenon in general relativity, in which the passing of a GW leaves a permanent, transverse displacement in a ring of test masses \citep[][]{1974zeldovich,demetrios,1992thorne,1992blanchet,2018talbot}. The observational consequences of GWM are plentiful, ranging from improved parameter estimation for individual compact object mergers \citep[][]{1987Braginsky,2009favata,2010Favata,2016Lasky,2021Tiwari}, the existence of astrophysical GWM backgrounds in various frequency bands \citep[][]{2022Zhao}, manifestations in cosmology \citep{2016Tolish,2017Bieri,2022niko}, and finally its influence on the radiation from the cosmic microwave background (CMB).

The latter is perhaps the most intriguing, as CMB observations constitute the cornerstone of our current understanding of the early Universe \citep{2020PlanckVI}. GWM results in both the red-shifting of CMB radiation as well as the mixing of power from different modes. Both of these aspects have been beautifully studied in \cite{2020Madison} and \cite{2021Madison} for the first time. However, the aforementioned works do not attempt to make an explicit connection with astrophysical populations of sources, opting instead to treat the problem with more generality. {Very recently, the former point has been addressed in \citet{2024Boybeyi}, showing how realistic populations of mergers produce an accumulation of GWM that affects the astrometry of far-away point sources. However, this study focus solely on the deflection to CMB photon paths as an observable consequence of GWM.}

In this work, we focus instead on the temperature fluctuations directly imprinted on the CMB by the GWM of massive black hole (MBH) mergers. We find that the accumulation of GWM caused by multiple mergers grows such fluctuations in a random-walk-like fashion \citep[see also][]{2024Boybeyi}, leading to a rich and potentially extremely informative phenomenology of patterns that {lives  hidden} orders of magnitude below the observational limitations of current all-sky CMB surveys.

\section{GWM from binary mergers sourcing CMB temperature fluctuations}
GWs propagate from their source radially as transverse-traceless waves. Here we assume the source to be a compact object binary with total mass $M$ and inclination $\iota$. We expect GWM to be present once information about the binary merger reaches a given space-time location, at a luminosity distance $D$. {The strain tensor for the non-linear GWM of a BH binary at merger is given by \citet{2009favata}}:
\begin{align}
\begin{split}
    h_{ij}^{\rm{mem}} &= -\frac{1}{576} \frac{GM}{c^2 D} \\ \nonumber & \times\left(\begin{array}{ccc}
    0 & 0 & 0 \\
    0 & (17+\cos^2\iota)\sin^2 \iota & 0 \\
    0 & 0 & -(17+\cos^2\iota)\sin^2 \iota
    \end{array}\right),
\end{split}
\end{align}
where $G$ is Newton's gravitational constant, $c$ is the speed of light, and the GW had been propagating in the $x$ direction. Note that here we assumed equal mass binary components, and will do so for the remainder of this work. Photons travelling through space may experience permanent distortions and deflections caused by the GWM from one such merger event. Here we focus solely on the resulting change in photon wavelength. While GWM also causes path deflections \citep[][see also \cite{1999kopeikin, 2011book} for the analogous effect from general GWs]{2020Madison}, the latter will be dominated by gravitational lensing from the entire dark matter halo, by a factor $\sim M_{\rm Halo}/M$. We leave the convolution of these two effects for future work.

We devise a strategy to calculate the imprint of GWM on 2D patches of the CMB\footnote{See \url{https://www.youtube.com/watch?v=s2d9ROsMoF0} for a visualisation of our setup.}. We model MBH mergers as events, distributed in a local box placed at a particular redshift. Each event $E$ is characterised by the following labels:
\begin{align}
  E:=  \left[t_{\rm m},x,y,z,M,\iota,\gamma \right],
\end{align}
where $t_{\rm m}$ is the local time of the merger, $x$, $y$ and $z$ its spatial coordinates, while $\gamma$ is an additional orientation angle of the MBH binary with respect to the observer reference frame. Each event is given a total mass $M$, and is sampled from a uniform distribution in time, inclination and orientation. The spatial coordinates of the events are instead sampled from a 3D distribution with a given power spectrum of the form:
\begin{align}
    P_{\rm 3D}(k) \propto k^{-n},
\end{align}
with $k$ being the Fourier mode. Large values of $n$ describe highly spatially clustered events. For the majority of our analysis, we draw events from an uniform distribution, i.e. in the following we set $n=0$ unless otherwise stated. Note that here we do not consider any possible clustering in time. \newline \newline
For our purposes it suffices to assume that GWM acts as a step function, which propagates information about the merger event outwards in a shell. We model it with a  Heaviside function ${\Theta_{\rm H}}$:
\begin{align}
    h_{ij}(\mathbf{x},t) = h_{ij}^{\rm {mem}} \times \Theta_{\rm H}\left(t-t_{\rm m} - \frac{|\mathbf{x}_\mathrm{m}-\mathbf{x}|}{c} \right),
\end{align}
where $\mathbf{x}_\mathrm{m}$ is the location of the merger. CMB photons that intersect the shell are therefore instantaneously affected by GWM. Their wavelength is either stretched or squeezed by the transverse-traceless metric strain projected along the photon propagation vector. The resulting change in wavelength is therefore maximised when photons intersect the expanding GWM shell at right angles. Across CMB patches, this effect manifests as a correlated wavelength perturbation pattern.  \newline \newline



Hereafter, we will use  Wien's displacement law to relate a small change in photon wavelength to a corresponding change in bulk blackbody temperature:
\begin{align}
    \frac{\Delta \lambda_{\rm ph}}{\lambda_{\rm ph}} \approx - \frac{\Delta T}{T},
\end{align}
where $\lambda_{\rm ph}$ is the photon's wavelength at peak spectral radiance and $T$ is the mean CMB temperature.
To produce 2D temperature fluctuation maps, we can apply the GWM of single or multiple events to a surface of photons, being careful to project each GWM strain correctly.


\section{Temperature fluctuations from individual Major Mergers}
\label{sec:resindividual}

\begin{figure}
    \centering
    \includegraphics[width = 0.45\textwidth]{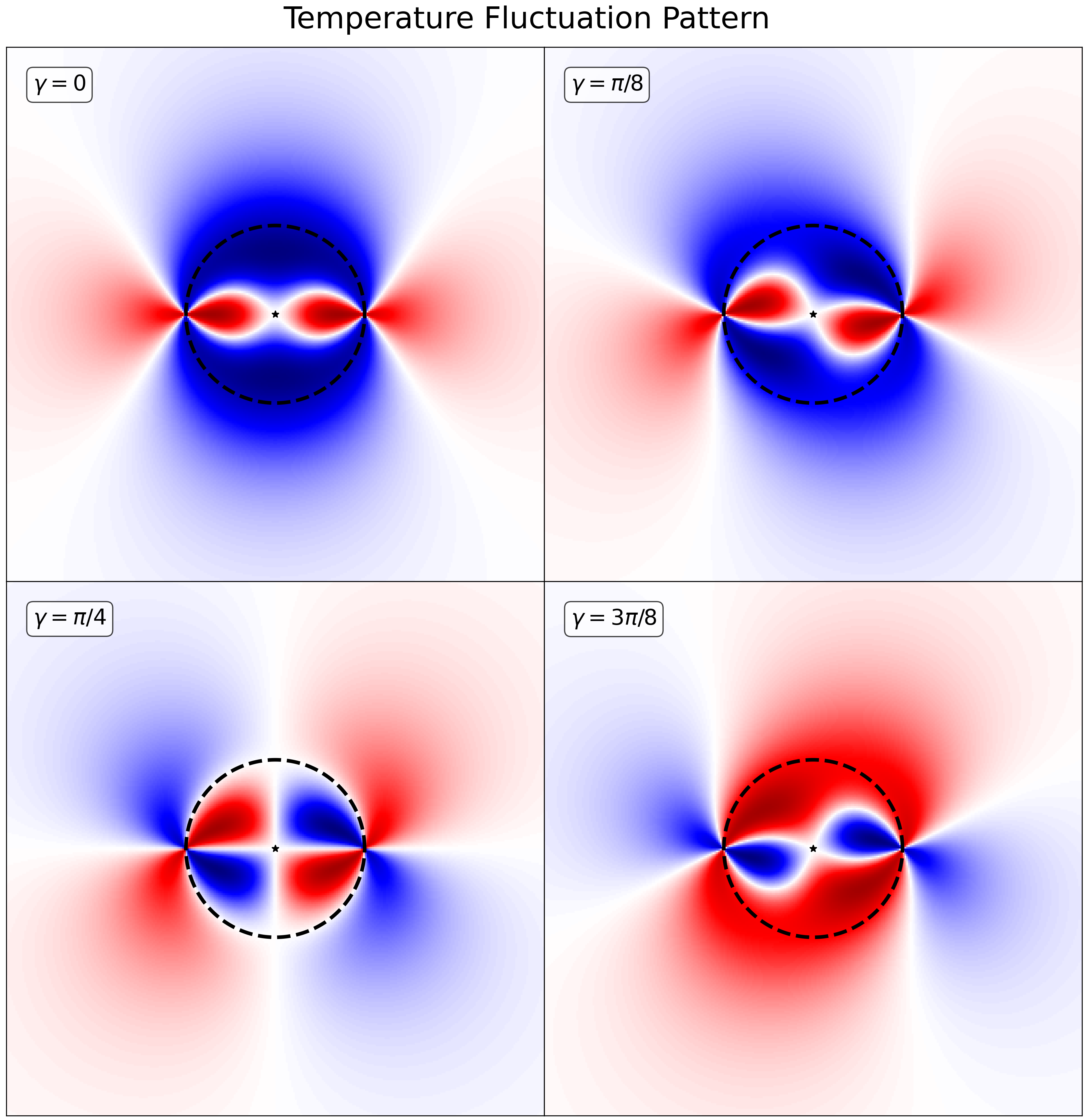}
    \caption{Temperature fluctuation patterns {for photons that have exited a local box, caused by} an individual merger event with different orientation angles $\gamma = 0$. The amplitude and physical scale of the fluctuations can be found by using Eq. \ref{SingleScaling}. Note that the average temperature fluctuation always vanishes, by merit of the tracelessness of the GWM tensor. {The projected scale of the patterns is related to the size of the GWM shell at the time the surface of photons crosses the coordinates of the source}, denoted by the grey dashed line.}
    \label{fig:TFLUC}
   
\end{figure}
\begin{figure*}
    \centering
    \includegraphics[width = 1\textwidth]{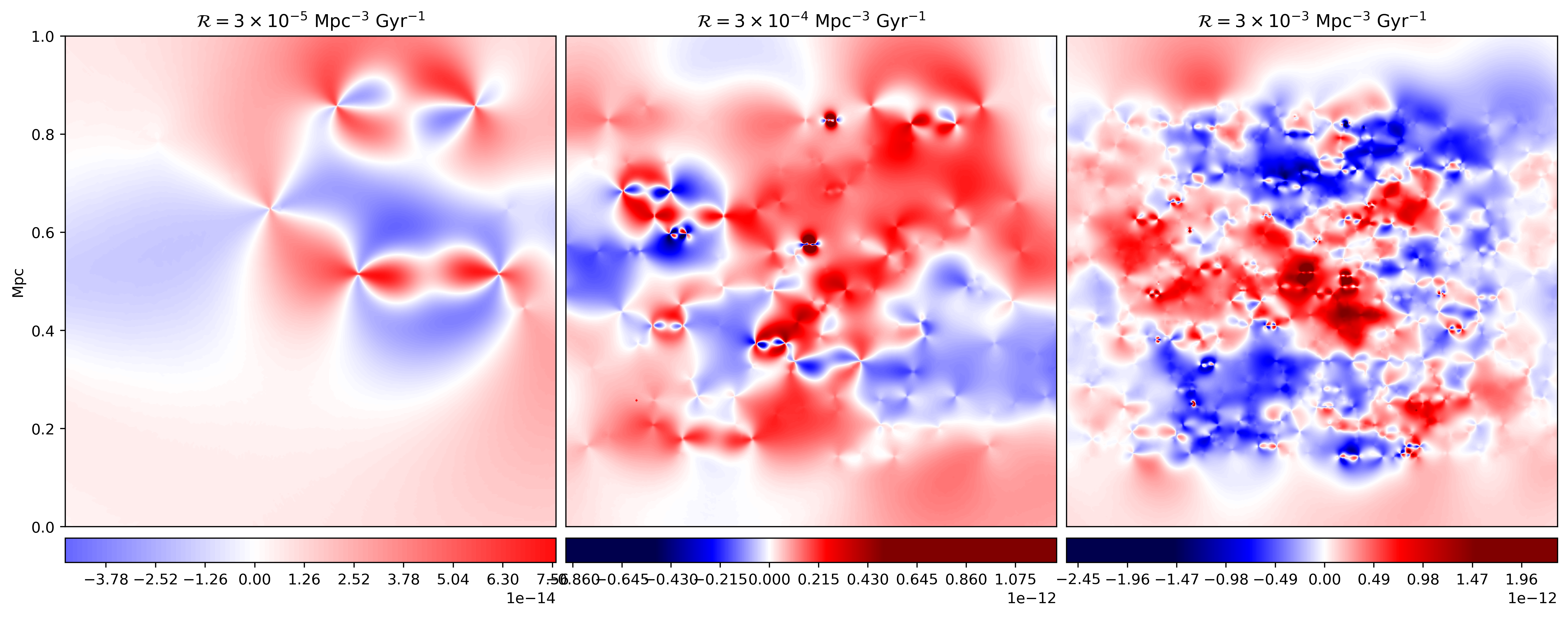}
    \caption{Three realisations of the CMB temperature fluctuations caused by the GWM of a population of mergers. The merger events have a mass of $M=10^9$ M$_{\odot}$ and are drawn from a uniform distribution in space-time and inclination and orientation angles. As the number of mergers is increased according to a specified volumetric merger rate $\mathcal{R}$, both the amplitude and the standard deviation of the temperature fluctuations grows following a random walk scaling. The colorbars are adjusted to account for this behaviour. With sufficient sources within a box ($\sim 10$ or more), the temperature fluctuations tend to follow universal power spectrum with a scaling $\propto k^{-2.7}$, where $k$ are the modes.}
    \label{fig:tflucpop}
\end{figure*}
In Fig. \ref{fig:TFLUC} we showcase the full GMW imprint of an individual merger event on the CMB temperature, where we vary both inclination and orientation angles. We find that the shape of the patterns is determined by the angles $\iota$ and $\gamma$, while the amplitude and scale are determined by the mass $M$, the merger time $t_{\rm m}$, the location of the merger along the line of sight and the inclination (to a lesser degree). As expected, the average temperature across the 2D image is  unaffected. The patterns roughly scale with the size of the GWM shell (grey dashed line) at the time the photon surface matches with the location of the source.
\newline \newline

We find the following scaling for the maximum amplitude of temperature fluctuations caused by an individual event:
\begin{equation}
        \frac{\Delta T_{\mathrm{max}}}{T(z_{\rm m})} = 1.72\times 10^{-15} \left(\frac{10^8 \mathrm{pc}}{R}\right)\left(\frac{M}{10^9 M_\odot}\right)\sin^2{\iota}
\label{SingleScaling}
\end{equation}
where $R > 0$ is the radius of the GWM shell and $T(z_{\rm m})$ denotes the mean CMB temperature at the redshift of the merger event. Here we scale our results with values typical for SMBH binaries, as they are the  loudest astrophysical source of GWs, and therefore GWM. The maximum temperature perturbation scales inversely with $R$, meaning that photons that intersect the GWM shell close to the merger experience the largest temperature perturbations.

In order to turn a physical scale to an angular scale we need to specify a redshift. Here we scale our results with a redshift $z=1$, which roughly corresponds to the peak of the massive galaxy volumetric merger rate \citep{2010Fakhouri,2021oleary,2022conselice}. We find the following scaling:
\begin{align}
    \frac{\Delta T_{\mathrm{max}}}{T(z_{\rm m})} \approx 3.54\times 10^{-13} \left(\frac{1'}{\beta}\right)\left(\frac{M}{10^9 M_\odot}\right) \left(\frac{\mathcal{D}_{\theta}(1)}{\mathcal{D}_{\theta}(z_{\rm m})} \right),
    \label{eq:betascaling}
\end{align}
where $\beta$ is the angular scale of the GWM shell, and we now neglect the inclination dependance. Here $\mathcal{D}_{\theta}$ is the angular diameter distance of a given cosmology, which peaks at roughly $z\sim1$ in $\Lambda$-CDM \citep{1993peebles}. For our purposes, redshifts of $\sim 1$ are therefore the most unfavourable, as by this effect alone temperature fluctuations at a given angular scale will have a factor $\sim 2$ higher amplitudes if produced at $z=7$ rather than $z=1$. Note that here we chose to scale the results with one minute of arc, as the typical resolution of state-of-the-art all-sky CMB missions is $\sim 5'$ \citep{2011core,2020Planck}.
\newline \newline
In reality, the temperature patterns produced by the GWM of mergers are not static. Different isochrone surfaces of CMB photons continuously traverse  any local box that contains MBH mergers, causing the projected patterns of temperature fluctuations t vary in time, roughly in proportion to their physical scale (as $\sim R/c$). We showcase several movies of time-evolving fluctuations, the footage of which can be found here \footnote{ \url{https://www.youtube.com/watch?v=2iwVHoZ6VF4&list=PLllD9CwiToPKY7hjYHaJDGrB5KBjRPyi4&index=1}}. For a single event, the GWM shell propagates radially with the passing of time, affecting photons at greater distances from the source. The result is to enlarge the physical scale of the pattern as well as decreasing its amplitude. We can express the typical temperature fluctuation, as seen by a local observer, as a function of the elapsed time $\tau$ after the merger has occurred:
\begin{align}
    \frac{\Delta T_{\mathrm{max}}}{T(z_{\rm m})} =& 5.56\times 10^{-15} \left(\frac{10^8 \mathrm{yr}}{\tau}\right)\left(\frac{M}{10^9 M_\odot}\right).
\end{align}


\section{Temperature Fluctuations from populations of Mergers}
Fig. \ref{fig:tflucpop} shows three realisations of the temperature fluctuation patterns in a local box of size $\sim (100\, {\rm{Mpc}})^{3}$, populated with $10$, $100$ and $1000$ sources, respectively. Here the events all have the same mass of $10^{9}$ $M_{\odot}$, but are otherwise drawn from uniform distributions. While the resulting temperature fluctuations seem to increase in complexity and resolution, they are composed of superpositions (ignoring non-linearities of the order $h_{ij}^2$) of the individual patterns showcased in Fig. \ref{fig:TFLUC}, appearing at different scales and with different orientations and inclinations.

\begin{figure}
    \centering
\includegraphics[width=\columnwidth]{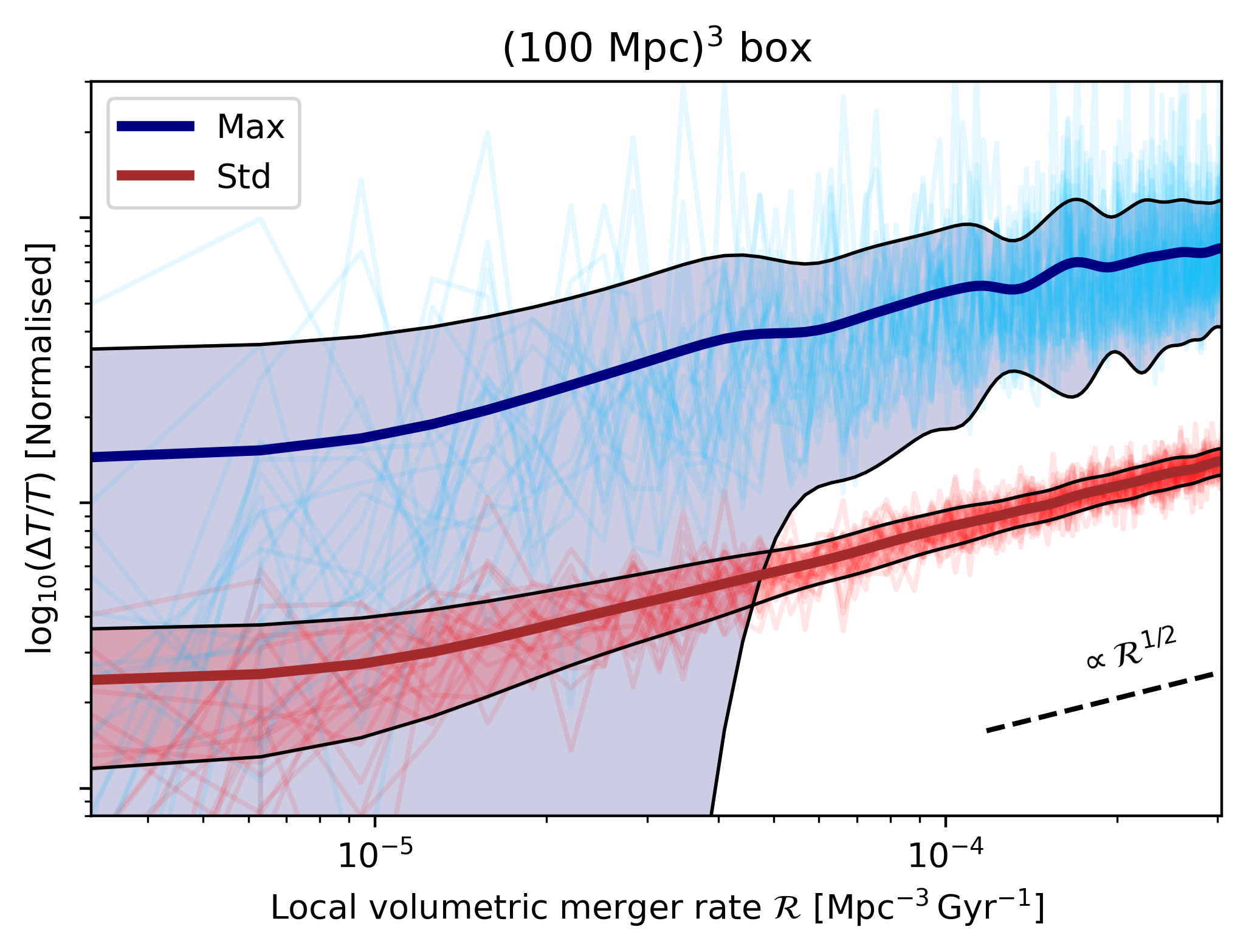}
    \caption{We show the maximum and the standard deviation of temperature fluctuations for CMB photons that passed through a (100 Mpc)$^3$ cosmological box. The box is populated with $10^{9}$ M$_{\odot}$ mergers according to a given volumetric merger rate, which are otherwise sampled uniformly in space, time, inclination and orientation. The results are averaged over a total of $\sim 3\times 10^4$ realisations (pale lines), an both the mean (solid lines) and the standard deviation (shaded areas enclosed by black lines) of the realisations are shown.}
    \label{fig:popscaling}
\end{figure}

As the mean of the fluctuations is always vanishing, we expect both the maximum and the standard deviation of temperature fluctuations within a given box to roughly follow a random walk process \citep[see also][]{2024Boybeyi}. Both of the aforementioned quantities should scale as $\sqrt{N}$, where $N$ is the total amount of merger events in the box. Here we explicitly confirm the hypothesis by setting up local boxes of various sizes. We populate the boxes with events corresponding to a given volumetric merger rate $\mathcal{R}$. A specific realisation of one such test is visualised in Fig. \ref{fig:popscaling}, showing both the maximum and standard deviations of the temperature fluctuations averaged over a total of $\sim 3\times 10^4$ realisations with varying number of events. These particular realisations were performed in a box of length 100 Mpc, roughly corresponding to a few degrees of angular scale at $z=1$. The masses are set to $10^9$ M$_{\odot}$ and the remaining parameters are drawn from uniform distributions. Throughout our tests, we find that both the maximum and the standard deviation of temperature fluctuations in a given CMB patch scale with the square root of the average local volumetric merger rate in the corresponding box. The maximum follows:
\begin{align}
    \frac{\Delta T_{\rm{max}}}{T(z_{\rm box})} = 7.28\times 10^{-13}\left(\frac{\mathcal{R}(M)}{10^{-4} \, \rm{Mpc}^{-3}\, \rm{yr}^{-1} }\right)^{1/2}\left(\frac{M}{10^{9}\, \rm{M}_{\odot}}\right),
    \label{eq:1}
\end{align}
though always presents a significant scatter, varying by a factor of approximately $2$ to $5$ at $1\sigma$, depending on the total number of events. Note that here $T(z_{\rm box})$ denotes the average local CMB temperature, a quantity that is well defined only for sufficiently small boxes. The standard deviation follows:
\begin{align}
    \frac{\Delta T_{\rm{std}}}{T(z_{\rm box})} = 1.05\times 10^{-14}\left(\frac{\mathcal{R}(M)}{10^{-4} \, \rm{Mpc}^{-3}\, \rm{yr}^{-1} }\right)^{1/2}\left(\frac{M}{10^{9}\, \rm{M}_{\odot}}\right),
    \label{eq:2}
\end{align}
with significantly less scatter. Volumetric merger rates of order $10^{-4}$ Mpc$^{-3}$ Gyr$^{-1}$ to $10^{-3}$ Mpc$^{-3}$ Gyr$^{-1}$ are what is typically observed for the major mergers of massive galaxies with stellar mass above $10^{11}$ M$_{\odot}$ at $z\sim 1$ \citep{2006bell,2021oleary}. According to observations, such galaxies typically host black holes in the range $10^{7}$ M$_{\odot}$ to several $10^{9}$ M$_{\odot}$ depending on galaxy morphology \citep{2015reines,2021zhu}, with an occupation fraction that approaches unity for more massive galaxies \citep{2016volonteri}. We scaled Eqs. \ref{eq:1} and \ref{eq:2} with the volumetric merger rates and BH masses appropriate to such galaxies, assuming that the BH merger rate would be roughly of the same order. As discussed before, the temperature fluctuations vary in time. We showcase an animation for the case of 20 randomly distributed sources here \footnote{ \url{https://www.youtube.com/watch?v=jli4n9j6BDk&list=PLllD9CwiToPKY7hjYHaJDGrB5KBjRPyi4&index=3}}. We observe the superposition of patterns from individual sources, each of which expands and fades during its evolution according to timescales of order $R_i/c$, where $R_i$ are the scales of each individual pattern. \newline \newline

\begin{figure}
    \centering
    \includegraphics[width=\columnwidth]{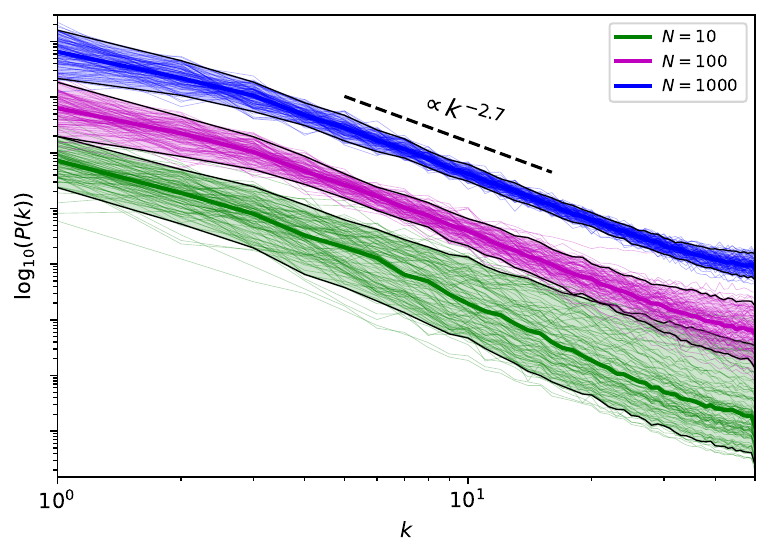}
    \caption{Realisations of the power spectrum of CMB temperature fluctuations caused by the GWM of $10$, $100$ and $1000$ mergers wihtin a local $(30 \, \rm{ Mpc})^3$ box (pale lines). We find that the power spectra follow a universal scaling $\propto k^{-2.7}$, where $k$ are the modes of the image (here in units of image resolution). The mean (thick lines) and the 2$\sigma$ contours (shaded areas) are also displayed.}
    \label{fig:power}
\end{figure}

We find that the GWM effect of a population of mergers gives rise to a characteristic power spectrum in the resulting CMB temperature fluctuation. Fig. \ref{fig:power} shows the power spectrum from 100 merger population realisations of boxes of local size $(30 \, \rm{ Mpc})^3$ containing $10$, $100$ and $1000$ mergers. Due to the $\sqrt{N}$ proportionality of the maximum temperature fluctuation as described above, the amplitude of the power spectrum is similarly increasing with the density of mergers. The larger scatter present for lower merger density is caused by small-number statistics, where with too few sources the initial merger distribution can be very inhomogenous when drawn randomly. We fit a power law $P(k)\propto k^{n}$, to the curves displayed in Fig. \ref{fig:power}, disregarding the lowest ($k<3$) and highest modes ($k>40$) to avoid boundary and  resolution effects of the temperature map. Here $k$ is in units of the image resolution, and can be easily transformed to physical scales given the scale of the CMB patch. We find the best fit scaling to be $n = -3.07\pm0.62$, $n = -2.82\pm0.30$, $n = -2.71\pm0.11$ for patterns caused by 10, 100 and 1000 sources respectively. 
We find that the scaling of the power spectrum of temperature fluctuations converges  to a universal value of:
\begin{align}
    &P^{\rm{GWM}}_{\rm{2D}}(k)\propto k^{-2.7},
\end{align}
within all CMB patches that contain a sufficient number of merger events.

\section{Towards Detecting GWM imprints}
The temperature fluctuations patterns analysed in this work showcase how every single BH merger leaves its signature on CMB radiation. Though hidden below a myriad of other signals \citep[see e.g.][]{2020planckI,2022ApJ...928...16R,2023PhRvD.107d3521B}, the entire merger history of BHs is marked on the oldest image of our Universe. Nevertheless, while contemplating such aesthetically pleasing thoughts, the astrophysicist is called to ask the following question: are such patterns actually detectable? Here we offer a few insights, in the understanding that detecting such fluctuations may well be exceedingly challenging: Though real, they are small in amplitude and have to be de-convolved from lensing, path deflections and other sources of noise. We welcome input from the broader community to explore some non trivial detection techniques \citep[see e.g.][for an attempt to reduce contamination from astrophysical noise sources]{2023kusiak}.
\newline \newline
\textsc{Holmberg 15} is a nearby ultra-massive BH with a mass of $\sim 4\times 10^{10}$ M$_{\odot}$, located at a redshift of $z\sim 0.05$ \citep{2019holmberg}. Is it possible to exclude that it experienced a major merger in its recent past? Suppose a CMB microwave interferometer sensitive to wavelengths $\lambda_{\rm{sens}}$ of $\sim$ 2 mm and temperature fluctuations of $\Delta T _{\rm sens}$ \citep[see][as an example]{2004Dickinson}. The latter would require the following baseline $B$ in order to resolve the GWM pattern around a presumed major merger with total mass $M$ at $z=0.05$:
\begin{align}
    B \approx 22.1\, \rm{km} \times \left(\frac{\Delta T _{\rm sens}}{10^{-5}\, \rm{K}}\right)\left(\frac{4\times 10^{10}\, \rm{M}_{\odot}}{M}\right) \left(\frac{2 \, \rm{mm}}{\lambda_{\rm{sens}}}\right),
\end{align}
where we used Eq. \ref{eq:betascaling} and $T_{\rm CMB}=2.73$ K. {While beyond the size of many current telescope arrays}, such baselines are clearly realistic. { However, at the corresponding resolution of tens of milli-arcseconds the GWM pattern would appear at scales of hundreds of pc, only approximately $10^2$ years after merger}. Improving the temperature fluctuation sensitivity beyond the typical values of current all-sky CMB surveys would allow to measure the patterns at larger scales, and probe whether a merger occurred further back in time. It is worth noting here that the GWM fluctuations will most likely be self-lensed, increasing their amplitude. \newline \newline
Our estimates for the random walk like accumulation of GWM apply to single mass populations of sources in local boxes. A complete assessment requires modelling the merger rates of all MBHs across cosmic time \citep[see][for a possible approach]{2024Boybeyi}. At higher redshifts in particular, however, the mass spectrum and merger rates of MBHs are purely speculative. The latter may well be much higher than what is adopted in this paper \citep[see some estimates on the merger rate of primordial black holes for an extreme example][]{2017aliha,2024huang}. We have shown that the power spectrum of temperature fluctuations is universal as soon as sufficient mergers are present in a given patch of CMB. For a volumetric merger rate of $10^{-3}$ Mpc$^{-3}$ Gyr$^{-1}$, this is realised for a box with angular scale of approximately $1^{\circ}$, of which there are 41253 independent realisations on the full sky. Knowing the scaling a priori, is it possible to coherently add up power from all such patches in a way that suppresses other contributions? Additionally, among the 41253 there are between 2 and 3 patches that are 4 sigma statistical outliers. Here outlier could refer to several of the quantities relevant to the accumulation of GWM, e.g. the merger rate, typical MBH masses, or simply to the realisation statistics of the random walk. There is always some probability that in some region of the sky the effects we discussed may be unusually large, though quantifying this would require a fully cosmological treatment over the entire sky. In the same vein, we note that our analysis should be convolved with an evolving cosmological metric, which will most likely increase the magnitude of the effect by an appreciable amount.
\section*{Data availability}
All animations are stored and available on Zenodo, under the following \href{ https://zenodo.org/records/14009709?token=eyJhbGciOiJIUzUxMiJ9.eyJpZCI6IjhlMWY5MzhiLTU2MDAtNGYzMS04N2YzLTA4NDVmMWRiMGJiNyIsImRhdGEiOnt9LCJyYW5kb20iOiI4NGMyNDIyZWQ2NzJlOTIwYzdlY2RiYzU3NjMxYTA4ZiJ9.eMpNREpYUHYyQqtdRblljNl1c4kyvOFXRSZmclpceBo0wya-QlXobiW9S5pep0AeR1sDrn7hnohfVoczw6q7_g}{repository}.
\begin{acknowledgements}
L.Z. and K.H. acknowledge support from  ERC Starting Grant No. 121817–BlackHoleMergs. M.M.T. acknowledges support from the Ministerio de Ciencia, Innovación y Universidades del Gobierno de España through the “Ayuda para la Formación de Profesorado Universitario" (FPU) No.~FPU19/01750 and the "Ayuda FPU Complementaria de Movilidad para Estancias Breves en Centros Extranjeros"  No.~EST23/00420, and from the Spanish Agencia Estatal de Investigación (grant PID2021-125485NB-C21) funded by MCIN/AEI/10.13039/501100011033 and ERDF A way of making Europe. D.ON gratefully acknowledges support from Villum Fonden Grant No. 29466 and the Danish Independent Research Fund through Sapere Aude Starting Grant No. 121587. P.K. acknowledges support from the Stefan Rozental og Hanna Kobylinski Rozentals Fond. The authors acknowledge Daniel J. D'Orazio, Johan Samsing and Martin Pessah for helpful discussions. The authors are grateful for the existence and convenient location of Søernes Ølbar.
\end{acknowledgements}

%
%

\bibliographystyle{aa} 
\bibliography{aa}

\end{document}